\documentstyle[aps,eqsecnum,graphicx,floats]{revtex}
\begin{document}
\draft
\author{C. Cabrillo\cite{email}, J. L. Rold{\'a}n, P. Garc{\'\i}a-Fern{\'a}ndez}
\address{Instituto de Estructura de la Materia, CSIC,
Serrano 123, 28006 Madrid, Spain.}

\date{\today}
         
\title{Quantum noise reduction in singly resonant optical devices}

\maketitle

\begin{abstract}
Quantum noise in a model of singly resonant frequency doubling 
including phase mismatch and driving in the harmonic mode 
is analyzed. The general formulae about the fixed points and 
their stability as well as the squeezing spectra calculated 
linearizing around such points are given. The use of a nonlinear 
normalization allows to disentangle in the spectra the dynamic
response of the system from the contributions of the various
noisy inputs. A general ``reference'' model for one-mode systems is 
developed in which the dynamic aspects of the problem are not contaminated
by static contributions from the noisy inputs. The physical insight gained 
permits the elaboration of general criteria to optimize the noise 
suppression performance.  With respect to the 
squeezing in the fundamental mode the optimum working point 
is located near the first turning point of the dispersive bistability induced by 
cascading of the second order nonlinear response. The nonlinearities 
induced by conventional crystals appear enough to reach it being the 
squeezing ultimately limited by the escape efficiency of the cavity.
In the case of the harmonic mode both, finite phase mismatch and/or  
harmonic mode driving allow for an optimum dynamic response of the 
system something not possible in the standard phase matched 
Second Harmonic Generation. The squeezing is then limited by the 
losses in the harmonic mode, allowing for very high degrees of 
squeezing because of the non-resonant nature of the mode.
This opens the possibility of very high performances using artificial 
materials with resonantly enhanced nonlinearities. It is also shown how 
it is possible to substantially increase the noise reduction 
and at the same time to more than double the output power for 
parameters corresponding to reported experiments.
\end{abstract}
 

\section{Introduction}{\label{I}}
Second Harmonic Generation (SHG) has nowadays quite a long tradition as a mean of 
squeezed light generation \cite{Per88,Siz90,Kur93,Pas94,Ral95,Tsu95,You96}. 
The preferred experimental setup has been the doubly resonant configuration 
as, at least in principle, permits arbitrarily large squeezing. However, such 
scheme has been hampered by the technical difficulties arising from keeping
the resonance in both modes simultaneously. Thus, in spite of the development
of very ingenious stabilizing procedures \cite{Kur93}, 
for the moment it has been only possible to maintain the double
resonance for a few seconds. Certainly, this kind of experimental delicacy can 
hardly surprise when dealing with the generation of non-classical 
states of light. In view of such difficulties, some experimental efforts 
have been recently redirected to singly resonant configurations 
\cite{Pas94,Ral95,Tsu95}. Although, the maximum noise suppression is then
limited to a 90\% \cite{Pas94}, the efforts resulted in very stable
intense squeezed light sources with degrees of squeezing
even surpassing those reported in the doubly resonant counterparts 
\cite{Tsu95}. This evolution highlights the importance of reducing
to a minimum the technical demanding of new proposals in a so 
experimentally challenging field. 

At the same time, singly resonant Optical Parametric Oscillation (OPO), the 
most successful method to squeeze the vacuum \cite{Pol92,Bre95}, has been 
generalized to singly resonant Optical Parametric Amplification (OPA), i.e.\, 
a laser driving in the harmonic mode has been added, again showing an 
extraordinary stability at quite high noise suppression values in the 
fundamental mode \cite{Sch96a}. Although the squeezed beams are in 
this case much less intense than the in SHG counterpart, this setup permits 
a control of the phase of the squeezed quadrature, something which 
allowed a spectacular demonstration using quantum tomography, of the different 
kinds of squeezed states \cite{Bre97}. 

In view of this experimental success it seems timely to extend 
the quantum mechanical model beyond the pure phase matched cases.
More specifically, we address here quantum noise reduction in an 
extension of the conventional singly resonant SHG to include also 
a coherent input in the harmonic mode as well as phase mismatch 
between the interacting waves. 
  
On the other hand, an increasingly number of papers is being devoted 
to study quantum noise in systems combining different 
kinds of nonlinearities (see, for instance, 
\cite{Sun95,Mar95a,Mar95b,Cab97,Khe98} for some recent 
contributions). In particular, the combination of $\chi^{(2)}$ 
with Kerr-like $\chi^{(3)}$ nonlinearities in c.w.\ cavity systems 
has been quite extensively studied 
\cite{Cab97,Tom84,Tom86,Gar89,Cab92a,Cab92b,Cab93,Kry96,Khe97}. 
Even exact full quantum results have been obtained showing, for instance, 
the emergence of tristability not present 
in the classical counterpart \cite{Khe97}. 
With respect to the squeezing performance the results appear as 
very promising  at least in degenerate doubly 
resonant configurations \cite{Cab97,Cab93}. The simplest system 
from the implementation point of view, combining this two kinds of 
nonlinearities that the authors can think of is precisely a singly 
resonant second order nonlinear system with phase mismatched 
interacting waves as then, by virtue of the cascading effect, an 
effective Kerr-like third order nonlinearity appears.

In order to make the search of strong noise reduction through the 
parameter space affordable we are bounded to the standard linearization 
procedures, the only capable of yielding analytical results. Inside 
the linear approximation perfect squeezing is possible at dynamic 
instabilities. We make use of this fact to find optimum working 
points showing up maximum squeezing. They are, however, an artifact 
of the method as the linear approximation breaks down at the 
instabilities. What matters for the practical implementation  
of new squeezed light sources (our ultimate goal) are the optimum 
paths through the parameter space approaching such points. Let us 
explain a little more what we mean. Fixed a particular parameter 
there will be a set of values of the remaining parameters (including 
the frequency as such) tuned up to yield the maximum noise reduction. 
When the chosen parameter is varied an optimum path is defined by the 
set of parameter values maximizing the noise reduction at any stage 
of the variation. The real optimum working point will be somewhere 
along these paths before reaching an instability. 
Thus, these paths would guide the experimentalist 
towards the optimum working point in the real experimental setup.
The essence of our approach to find such paths will be to isolate 
the dynamic aspect of the squeezing behavior from the 
static contributions to the noise coming from the different inputs.
A simple adequate normalization will disentangle the two aspects of 
the quantum noise behavior. This simplifies the analysis 
sufficiently to allow a characterization of the optimum paths.

Another crucial issue when dealing with the squeezing performance of a 
system is precisely the election of the most relevant parameter to compare the 
different configurations with each other or with the reported related 
experiments. There is no universal criterion to determine the 
squeezing efficiency of a given device. An efficient setup
regarding power consumption, i.e., when compared for fixed input power 
could well be deceptive when compared for the same output power and 
perhaps inadequate to some spectroscopic applications. 
However, within the state of the art of the present squeezed light 
generators, the main concern is to improve the squeezing 
figures themselves having other considerations such as the power 
consumption a relative importance.  Under this perspective probably the 
parameter of utmost importance as far as c.w.\ resonant systems are concerned is 
the energy load inside the cavity. Indeed, the usual causes of squeezing 
degradation such as blue-light-induced red absorption come from an 
excessive mean photon number inside the cavity capable of 
significantly degrade the material optical response at the relevant 
frequencies. These considerations will lead us to define another normalization
this time useful for the evaluation of the squeezing efficiency with 
respect to the intra-cavity photon number.

The sketch of the article is as follows. In section \ref{QMM} the 
quantum mechanical model is presented. In section \ref{LEE} the 
evolution equations are linearized, the fixed points of the system 
obtained and their stability studied. Section \ref{SS} gives all the 
formulae regarding quantum noise spectra in the system. In section 
\ref{SP} a general approach to one-mode systems is developed which 
allows the definition of general criteria to characterize the optimum 
paths and applied to the specific case here addressed. 
Finally the limits of the model and possible implementations 
are thoroughly discussed in section \ref{DC}, concluding the article 
with a summary of the most relevant results obtained. 

\section{Quantum mechanical model}{\label{QMM}}
The system we want to address consists in a second order nonlinear 
medium coupling two modes of frequency $\omega$ (fundamental) and $2 
\omega$ (harmonic) respectively and placed inside a ring cavity resonant 
only with the fundamental mode. We will also assume just one input-output 
mirror of finite reflectivity.
The effect of phase mismatch when only the fundamental mode is driven 
has been experimentally studied in \cite{Whi96a} where bistability 
induced by cascading was demonstrated. The classical 
evolution equation of the fundamental mode, $\alpha$, as given in 
\cite{Whi96a}, reads
\begin{equation}
\label{eq:claeqa}
\frac{d\alpha}{dt} = -\left [ \gamma + i\delta + \nu K(\Delta k)
|\alpha|^{2} \right ]\alpha + \sqrt{2 \gamma_{c}}\, \alpha_{in} \,.
\end{equation}

\noindent
The nonlinear coupling depends on the wave vector mismatch 
$\Delta k = k(2\omega)-k(\omega)$ as $K(\Delta k) = 
2 \int_{0}^{L_{m}} \int_{0}^{z} u^{*}(\Delta k,z) u(\Delta k, z^{\prime}) 
d z^{\prime} d z / L_{m}^{2}$ being $L_{m}$ the length of the nonlinear 
medium, $u(k,z)$ the spatial dependence of the resonator mode and $\nu$ is 
proportional to the second order nonlinear susceptibility (see below).
Splitting $K$ in its real and imaginary parts, Eq.\ (\ref{eq:claeqa}) can
be recast as
\begin{equation}
\label{eq:claeqa2}
\frac{d\alpha}{dt} = 
-\left [ \gamma + \mu |\alpha|^{2} +
i(\delta+\Gamma |\alpha|^{2}) \right ]\alpha + 
\sqrt{2 \gamma_{c}} \alpha_{in} \,.
\end{equation}

\begin{figure}
\centerline{\includegraphics*[scale=0.5]{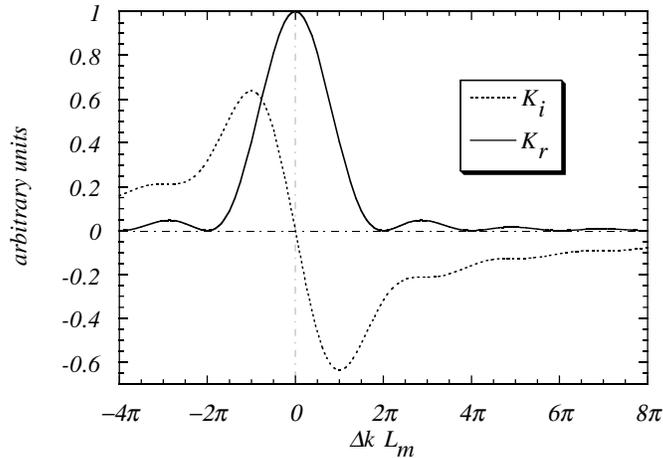}}
\caption{The dependence of $K_{i}$ and $K_{r}$ with respect to the phase 
mismatch.}
\label{fg:fg}
\end{figure}

\noindent 
For a plane wave geometry 
\begin{mathletters}
\label{eq:mu&Ga}
\begin{eqnarray}
\label{eq:mu}
\mu\;\equiv\;\nu K_{r} & = &\nu 
\left ( {\rm sinc}\frac{\Delta k L_{m}} {2} \right )^{2} \\
\label{eq:Ga}
\Gamma\;\equiv\;\nu K_{i} & = &\frac{2\nu} {\Delta k L_{m}} 
\left[{\rm sinc} \frac{\Delta k L_{m}} {2}
\cos \frac{\Delta k L_{m}} {2} -1 \right]\,,
\end{eqnarray}
\end{mathletters}

\noindent
where $K_{r}$ and $K_{i}$ denote the real and imaginary part of $K(\Delta k)$.
In this way the nonlinear dynamics is divided in a nonlinear 
absorption (the up conversion of photons) and a nonlinear dispersion
(the cascading effect). The behavior of both parameters with 
$\Delta k$ is plotted in Fig.\ \ref{fg:fg}. Notice that for any 
finite $\Delta k$ the nonlinear dispersion is also finite, periodically 
completely dominating (null frequency doubling).

Quantization of Eq.\ (\ref{eq:claeqa2}) is then
accomplished independently for each effect. Regarding the nonlinear
absorption we use the two-photon model proposed in \cite{Col91}, while
nonlinear dispersion is accounted for by a fourth order
Hamiltonian, $H = (\hbar \Gamma/2)\, a^{\dagger \,2} a^{2}$ as in the
standard theory of optical Kerr effect. It represents a
Hamiltonian modification of the two-photon absorption model so that the
quantum mechanical equation reads
\begin{equation}
\label{eq:quant}
\frac{d a}{dt} = 
-\left [ \gamma + i \delta + (\mu + i \Gamma)a^{\dagger}a \right ]a + 
2\sqrt{\mu}\,a^{\dagger}b_{in}+\sqrt{2 \gamma_{c}}\, a_{in} +
\sqrt{2\gamma_{s}}\,w_{in}\,,
\end{equation}

\noindent
where Latin characters denote the annihilation operators for the
corresponding classical (Greek characters) modes.
Two extra terms not present in the classical analog appear, namely, a
white noise input, $w_{in}$, accounting for the fluctuations induced 
by the scattering and the absorption in the crystal 
($\gamma_{s} = \gamma-\gamma_{c}$) and a parametric ``gain'' 
term coming from the, classically empty, incoming harmonic mode, $b_{in}$. 
Eq.\ (\ref{eq:quant}) is complemented with the boundary conditions 
\cite{Pas94}
\begin{mathletters}
\label{eq:boundary}
\begin{eqnarray}
\label{eq:boundarya}
a_{out} & = & \sqrt{2\gamma_c}\, a - a_{in} \,,\\
\label{eq:boundaryb}
b_{out} & = & \sqrt{\mu}\,a^{2} - b_{in} \,,
\end{eqnarray}
\end{mathletters}

\noindent
from which the output spectra can be computed. Input fields are assumed
to be in coherent states. In particular, allowing a coherent state
different from the vacuum for the incoming harmonic mode we generalize
the system to the case of driving both modes. In the case 
$\Gamma = \delta = 0$, the squeezing properties as well as the applicability 
to quantum nondemoliton measurements of this system have been studied in 
detail in \cite{Sch95}. 

The used definitions for the creation 
operators give the following relations with the usual experimental 
parameters (see appendix in \cite{Sch95}): the input and output powers 
are $P_{\omega,in/out} = \hbar \omega \langle a^{\dagger}_{in/out} 
a_{in/out} \rangle$ and $P_{2\omega,in/out}= \hbar 2 \omega \langle 
b^{\dagger}_{in/out} b_{in/out}\rangle$; the circulating power is $\hbar 
\omega \langle a^{\dagger} a \rangle / \tau$ being $\tau$ the 
round-trip time and the single-pass power-conversion efficiency (in 
W$^{-1}$) is $2 \tau^{2} \nu/ \hbar \omega$.

\section{Linearized evolution equations and linear stability 
analysis}{\label{LEE}}
 Defining fluctuation operators as
\begin{mathletters}
\label{eq:fluctope}
\begin{eqnarray}
a & = &\alpha +\delta a \,, 
\label{eq:fluctopea}\\ 
a_{in,out} & = & \alpha_{in,out} +\delta a_{in,out} \,, 
\label{eq:fluctopeain}\\ 
b_{in,out} & = & \beta_{in,out}+\delta b_{in,out} \,, 
\label{eq:fluctopebin}
\end{eqnarray}
\end{mathletters}

\noindent
a linearization of Eqs.\ (\ref{eq:quant}) and (\ref{eq:boundary}) yields
\begin{eqnarray}
\label{eq:linquant}
\frac{d\,\delta a}{dt} & = &  
-\left [ \gamma + i \delta + 2(\mu + i \Gamma)|\alpha|^{2} 
\right]\delta a + \left [ 2\sqrt{\mu}\beta_{in} - 
(\mu+i \Gamma)\alpha^{2}\right ] \delta a^{\dagger} \nonumber \\
& + & 2\sqrt{\mu}\,\alpha^{*}\delta b_{in}+
\sqrt{2 \gamma_{c}}\,\delta a_{in} + \sqrt{2\gamma_{s}}\,w_{in}\,,
\end{eqnarray}

\noindent
and
\begin{mathletters}
\label{eq:linboundary}
\begin{eqnarray}
\delta a_{out} & = & \sqrt{2 \gamma_{c}}\,\delta a - \delta a_{in}\,,
\label{eq:linboundarya} \\
\delta b_{out} & = & 2 \alpha \sqrt{\mu}\, \delta a - \delta b_{in}\,,
\label{eq:linboundaryb}
\end{eqnarray}
\end{mathletters}

\noindent
being $\alpha_{in,out},\beta_{in.out}$ the mean values of the 
corresponding input and output modes and $\alpha$ a 
stable fixed point of the classical counterpart of 
Eq.\ (\ref{eq:quant}), i.e.\,
\begin{equation}
\label{eq:class}
\frac{d \alpha}{dt} = 
-\left [ \gamma + i \delta + (\mu + i \Gamma) |\alpha|^{2} \right 
]\alpha + 
2\sqrt{\mu}\,\alpha^{*}\beta_{in}+\sqrt{2 \gamma_{c}}\,\alpha_{in}
\,.
\end{equation}

\noindent
Equating to zero the l.h.s.\ of Eq.\ (\ref{eq:class}) a ``state equation''
for the fixed points is obtained, namely,
\begin{equation}
\label{eq:state}
\alpha = \frac{\sqrt{2 \gamma_{c}}\left \{ \left[ \gamma+\mu n -i 
(\delta + \Gamma n) \right ] \alpha_{in}+ 2 \sqrt{\mu} \,\beta_{in}\,
\alpha_{in}^{*} \right \}}
{(\gamma + \mu n)^{2}+(\delta+\Gamma n)^{2}-4 \mu |\beta_{in}|^{2}}
\,,
\end{equation}

\noindent
with $n = |\alpha|^{2}$. Let be $\theta,\, \phi$ and $\varphi$ the 
phases of $\alpha,\, \alpha_{in}$ and $\beta_{in}$ respectively. 
Then, dividing both sides of Eq.\ (\ref{eq:state}) 
by $e^{i\varphi/2}$ 
\begin{eqnarray}
\label{eq:statea}
|\alpha|e^{i(\theta-\varphi/2)} \left [ (\gamma + \mu n)^{2}
+ (\delta+\Gamma n)^{2} - 4 \mu |\beta_{in}|^{2} \right ] \: = \: & & 
\nonumber \\
|\alpha_{in}| \sqrt{ 2 \gamma_{c}} \left [ \left ( \gamma + \mu n- i 
(\delta+\Gamma n) \right ) e^{i(\phi-\varphi/2)}+
2 \sqrt{\mu}\,|\beta_{in}|e^{-i(\phi-\varphi/2)} \right ] \,. & &
\end{eqnarray}

\noindent
Taking the squared modulus in both sides a quintic equation for $n$ is 
obtained
\begin{eqnarray}
\label{eq:neqn}
0 & = & n \left [ (\gamma + \mu n)^{2}
+ (\delta+\Gamma n)^{2} - 4 \mu |\beta_{in}|^{2} \right ]^{2} - 
2 \gamma_{c}|\alpha_{in}|^{2} \left\{ (\gamma + \mu n)^{2}
+ (\delta+\Gamma n)^{2} + 4 \mu |\beta_{in}|^{2} + \right . \nonumber \\
& & \left . 
4 \sqrt{\mu}\,|\beta_{in}| \left[ (\gamma+\mu n) \cos(2 \phi-\varphi)+
(\delta+\Gamma n) \sin(2 \phi -\varphi) \right] \right\} \,.
\end{eqnarray}

\noindent
The real and the imaginary part of Eq.\ ($\ref{eq:statea}$) determine the 
$\sin(\theta-\varphi/2)$ and $\cos(\theta-\varphi/2)$ as functions of 
the solutions of Eq.\ (\ref{eq:neqn})
\begin{mathletters}
\label{eq:phases}
\begin{eqnarray}
\cos(\theta-\varphi/2) & = & \frac{|\alpha_{in}|}{|\alpha|} 
\sqrt{2\gamma_{c}}\,
\frac{(\gamma + \mu n+2\sqrt{\mu}\,|\beta_{in}|)\cos 
(\phi-\varphi/2)+(\delta+\Gamma n) \sin (\phi-\varphi/2)}
{(\gamma+\mu n)^{2}+(\delta+\Gamma n)^{2}-4 \mu |\beta_{in}|^{2}} 
\label{eq:cos}
\\
\sin(\theta-\varphi/2) & = & \frac{|\alpha_{in}|}{|\alpha|} 
\sqrt{2 \gamma_{c}}\, 
\frac{(\gamma + \mu n-2\sqrt{\mu}\,|\beta_{in}|)\sin 
(\phi-\varphi/2)-(\delta+\Gamma n) \cos (\phi-\varphi/2)}
{(\gamma+\mu n)^{2}+(\delta+\Gamma n)^{2}-4 \mu |\beta_{in}|^{2}} \,.
\end{eqnarray}
\end{mathletters}
\label{eq.sin}

\noindent
Eq.\ (\ref{eq:neqn}) allows for numerical calculation of the fixed 
points given the input fields. But it can be interpreted also as a 
linear equation for $|\alpha_{in}|^{2}$, i.e.\,

\begin{equation}
\label{eq:ain}
2 \gamma_{c}|\alpha_{in}|^{2} = \frac{n \left [ (\gamma + \mu n)^{2}
+ (\delta+\Gamma n)^{2} - 4 \mu |\beta_{in}|^{2} \right ]^{2}}
{|\gamma + \mu n + 2 \sqrt{\mu} |\beta_{in}| 
e^{i (2 \phi-\varphi)}|^{2} + 
4 \sqrt{\mu}\,|\beta_{in}| (\delta+\Gamma n) \sin(2 \phi -\varphi)} \,.
\end{equation}

\noindent
The positive character of the r.h.s.\ is not always guaranteed and therefore 
not for every value of the parameters a real positive $n$ is 
possible. Notice, however, that in the cases in which this happens a 
simultaneous change of the sign of $\delta$ and $\Gamma$ yields a 
consistent set of parameter values. As we shall see, this fact will have 
useful consequences in regarding the analysis of the quantum noise 
behavior in the system.

The stability of the fixed points is governed by the real part of the 
eigenvalues of the drift matrix associated with the linearized 
evolution equation (\ref{eq:linquant}). Very simple algebra yields
\begin{equation}
\label{eq:eigen}
\lambda_{\pm}=-(\gamma+2 \mu \,n) \pm \sqrt{|(\mu+
i \Gamma)\alpha^{2}-2 \sqrt{\mu}\,\beta_{in}|^{2}-
(\delta+2\Gamma\,n)^{2}}\,.
\end{equation}

\noindent
Provided that the real part of both eigenvalues are negative the fixed 
point will be stable. With respect to the phase-matched SHG case ($\Gamma =0$ and
$\beta_{in}=0$, always stable), although both $\Gamma$ and $\delta$ alone
tend to stabilize the dynamics, in combination are able of destabilize
the system.  A finite $\beta_{in}$, on the other hand, can promote
instability depending on its relative phase with respect to $\alpha$, the case
$\theta - \varphi/2 =\pm \pi/2$ maximizing the effect. All of these new
instabilities, however, correspond to zero eigenvalues without a finite
imaginary part. In other words, contrary to the double resonant SHG
there is no Hopf bifurcation and, consequently, no selfpulsing solution.

\section{Squeezing spectra}{\label{SS}}

For a given quadrature of the electric field, 
$X_{\theta_{s}}^{out}(t) \equiv a_{out}(t) e ^{-i \theta_{s}}+ 
a_{out}^{\dagger}(t) e^{i\theta_{s}}$, the
squeezing spectrum is simply the noise spectrum of such
a quantity, i.e.\,
\begin{eqnarray}
\label{eq:Sw}
S(\omega) & = & C \int_{-\infty}^{\infty} \langle \delta
X_{\theta_{s}}^{out}(t) \delta X_{\theta_{s}}^{out}(t+\tau) \rangle \, 
e^{-i\omega \tau} d \tau \nonumber \\
& = & C \int_{-\infty}^{\infty}  \langle \delta
X_{\theta_{s}}^{out}(\omega) \delta X_{\theta_{s}}^{out}(-\omega^{\prime})
\rangle d \omega^{\prime} \,,
\end{eqnarray}

\noindent
being $C$ some normalization constant and the averages are assumed
stationary. As a function of the annihilation and creation operators 
Eq.\ (\ref{eq:Sw}) is rewritten as
\begin{equation}
\label{eq:Sw2}
S(\omega) = C \left [
\langle \delta a^{\dagger}_{out}(\omega) \delta a_{out}(-\omega)\rangle +
{\rm Re}\{\exp(-i2\theta_{s})
\langle \delta a_{out}(\omega) \delta a_{out}(-\omega) \rangle \} 
\right ]\,,
\end{equation}

\noindent
where use has been made of the stationarity of the average and Re 
denotes real part. From this expression it is evident that the noise is 
minimized and therefore the squeezing effect maximized for a quadrature 
phase such as
\begin{equation}
\label{eq:Sw3}
S(\omega) = C \left [
\langle \delta a^{\dagger}_{out}(\omega) \delta a_{out}(-\omega)\rangle -
\left |
\langle \delta a_{out}(\omega) \delta a_{out}(-\omega) \rangle
\right | \right ]\,,
\end{equation}

\noindent
corresponding to a phase
\begin{equation}
\label{eq:sqphase}
\theta_{m} = \frac{\nu(\omega) - \pi}{2} \,,
\end{equation}

\noindent
where $\nu(\omega)$ is the phase of 
$\langle \delta a_{out}(\omega) \delta a_{out}(-\omega) \rangle$. 
The spectrum of the conjugate quadrature (i.e.\, with a phase
$\nu(\omega)/2)$ corresponds to a plus sign in Eq.\ (\ref{eq:Sw3}) and
by virtue of the Heisenberg principle shows an excess noise above the
vacuum. Taking $C=1$
(corresponding to vacuum noise units) and splitting Eq.\ (\ref{eq:Sw3})
into a vacuum noise component plus a normally ordered part we finally
arrive to
\begin{equation}
\label{eq:Swf}
S_{-,+}(\omega) = 1 + 
\langle : \delta a^{\dagger}_{out}(\omega) \delta a_{out}(-\omega) :
\rangle \mp \left | 
\langle : \delta a_{out}(\omega) \delta a_{out}(-\omega) : \rangle
\right | \,,
\end{equation}

\noindent
for both the squeezing and the ``stretching'' spectra.
After tedious but simple algebra, the spectra of the fundamental 
and second harmonic modes can be written as
\begin{mathletters}
\label{eq:spectrum}
\begin{eqnarray}
S^{a}_{-,+}(\omega) & = & 1 + 4 \gamma_{c} |B| \frac{N_{-,+}}{D}\,,
\label{eq:spectruma} \\ 
S^{b}_{-,+}(\omega) & = & 1 + 8 \mu n |B| \frac{N_{-,+}}{D}\,,
\label{eq:spectrumb} 
\end{eqnarray}
\end{mathletters}

\noindent
where $B =2\sqrt{\mu} \,\beta_{in} -(\mu +i \Gamma) \alpha^{2}$ and
\begin{mathletters}
\label{eq:Sc}
\begin{eqnarray}
N_{-,+} & = & 2 |B| (\gamma + 2 \mu n) \mp \sqrt{\left [ 
(\gamma + 2 \mu n)^{2} - (\delta + 2 \Gamma n)^{2} + |B|^{2} +\omega^{2}
\right ]^{2} + 4 (\gamma + 2 \mu n)^{2}
(\delta + 2 \Gamma n)^{2}}\,,
\label{eq:ScN}\\
D & = & \left[(\gamma+2\mu n)^{2} + (\delta + 2 \Gamma n)^{2} -
|B|^{2} - \omega^{2} \right ]^{2} + 4 (\gamma+2\mu n)^{2} \omega^{2}\,.
\label{eq:ScD}
\end{eqnarray}
\end{mathletters}

\noindent
The correlations defining the squeezing phase $\nu(\omega)$ are given 
by
\begin{mathletters}
\label{eq:corr}
\begin{eqnarray}
\label{eq:corra}
\langle \delta a_{out}(\omega)\delta a_{out}(-\omega) \rangle & = &
4\gamma_{c} B \left [\omega^{2} + |B|^{2} +
(\gamma + 2 \mu n)^{2} - (\delta + 2 \Gamma n)^{2} + i \, 
2 (\gamma + 2 \mu n) (\delta + 2 \Gamma n)\right ]/D \,,\\
\label{eq:corrb}
\langle \delta b_{out}(\omega)\delta b_{out}(-\omega) \rangle & = &
8\mu\,\alpha^{2} B \left [\omega^{2} + |B|^{2} +
(\gamma + 2 \mu n)^{2} - (\delta + 2 \Gamma n)^{2} + i\,
2 (\gamma + 2 \mu n) (\delta + 2 \Gamma n)\right ]/ D \,.
\end{eqnarray}
\end{mathletters}

\noindent
The trigonometric equations for the corresponding phases are quite 
complicated and rather useless. However, an interesting
consequence can directly be drawn from Eqs.\ (\ref{eq:corr}), namely, 
for detuning-s such as $\delta+2 \Gamma n =0$ the phases 
are independent of $\omega$ equaling those of $B$ and $\alpha^{2} B$
respectively.  

\section{Squeezing performance}{\label{SP}}
In the previous sections we have developed the general raw formulae 
regarding quantum noise in the system. This section is devoted to the 
analysis of the quantum noise behavior implied by them an in 
particular to proceed with our program of finding optimum quantum 
noise reduction. However, before any specific assessment of the 
quantum noise performance we will elaborate a little more on the 
formulae (mainly by adequate normalizations) in order to gain 
physical insight and ease our task. As an aftermath we shall obtain 
general results going well beyond the specifics of the system addressed 
here.

\subsection{General results concerning one-mode systems}
Let us begin defining a nonlinear
and a total decay rate as $\gamma_{nl} \equiv 2 \mu n$ and 
$\gamma_{t} \equiv \gamma +\gamma_{nl}$ respectively. 
We shall scale the evolution with this total decay rate defining an 
dimensionless time $\tau \equiv \gamma_{t}\,t$.
In the spectra (\ref{eq:spectrum}) the only dependence on $\theta$ is 
through $B$ disappearing for $\beta_{in}=0$. 
It is also possible to restrict this dependence to such
a term directly in Eq.\ (\ref{eq:linquant}) and 
Eq.\ (\ref{eq:linboundary}) by means of appropriate phase shifts of the 
modes. All together account for
\begin{eqnarray}
\label{eq:normeq}
\frac{d\,\delta c}{d \tau} & = &  
-\left [1 + i \Delta \right]\delta c + 
\tilde{B}\,\delta c^{\dagger} \nonumber \\
& + & \sqrt{2 \tilde{\gamma}_{nl}}\;\delta d_{in} +
\sqrt{2 \tilde{\gamma}_{c}}\;\delta c_{in} +
\sqrt{2\tilde{\gamma}_{s}}\,s_{in}\,,
\end{eqnarray}

\noindent
where the tilde represents divided by $\gamma_{t}$, 
$\Delta = \tilde{\delta} + 2 \tilde{\Gamma} n $, $\tilde{B} =
2\sqrt{\tilde{\mu}}\,\delta_{in} - (\tilde{\mu}+i \tilde{\Gamma})n$ 
and the modes are redefined as  
\begin{mathletters}
\label{eq:nfields}
\begin{eqnarray}
c & \equiv &  a \, e ^{-i \theta}\,,
\label{eq:c}\\
c_{in,out} & \equiv &
\frac{a_{in,out}}{\sqrt{\gamma_{t}}} \, e ^{-i \theta}\,,  
\label{eq:cinout}\\
 d_{in,out} & \equiv &  
\frac{b_{in,out}}{\sqrt{\gamma_{t}}} \, e ^{-i 2 \theta}\,,
\label{eq:dinout}\\
s_{in} & \equiv & \frac{w_{in}}{\sqrt{\gamma_{t}}} \, e ^{-i \theta}\,. 
\label{eq:sin} 
\end{eqnarray}
\end{mathletters}

\noindent
In agreement with the previous notation $\delta_{in}$ denotes the mean
value of $d_{in}$. The boundary conditions of the new modes are
\begin{mathletters}
\label{eq:nlinboundary}
\begin{eqnarray}
\delta c_{out} & = & \sqrt{2 \tilde{\gamma}_{c}}\,\delta c - \delta c_{in}\,
\label{eq:nlinboundaryc} \\
\delta d_{out} & = & \sqrt{2 \tilde{\gamma}_{nl}}\, \delta c - \delta d_{in}\,. 
\label{eq:nlinboundaryd}
\end{eqnarray}
\end{mathletters}

\noindent
For coherent states, the correlations of the new input modes remain
as white noise but in the scaled time $\tau$. We shall refer to the 
previous formulae as the tilde normalization.

The evolution equation (\ref{eq:normeq}) encodes the dynamic response of 
the intracavity system to a series of noisy input channels ($\delta 
d_{in}$, $\delta c_{in}$ and $s_{in}$). Quantum Mechanical 
consistency, i.e., conservation of equal-time conmutators, imposes a
fluctuation-dissipation relation which under this normalization reads
\begin{equation}
\label{eq:add1}
\tilde{\gamma}_{nl}+\tilde{\gamma}_{c}+\tilde{\gamma}_{s}=1\,.
\end{equation}

\noindent
The evolution equation (\ref{eq:normeq}) along with Eqs.\ 
(\ref{eq:nlinboundary}) are now written in such a way that the input-output
couplings are real-valued as in the standard input-output formalism 
\cite{Coll84,Gar91}. This is a completely general result. Provided a well 
defined linearized theory in the sense of preserving equal-time 
conmutators we only need the adequate set of phase shifts of the input 
channels (a trivial unitary transformation preserving conmutators) making the 
couplings real-valued to obtain a theory formally equal to the standard 
input-output formalism simply because this is the theory preserving 
the equal-time conmutators when the couplings are real-valued. Thus, 
for any system with only one effective mode there is a formulation in which the 
intracavity field follows
\begin{equation}
\label{eq:geneqnorm}  
\frac{d\,\delta c}{d \tau}  =   
-\left [1 + i \Delta \right]\delta c + 
\tilde{B}\,\delta c^{\dagger} + \sum_{n=1}^{N} 
\sqrt{2 \tilde{\gamma}_{n}} \,\delta c_{in}^{n}\,,
\end{equation}
with
\begin{equation}
\label{eq:genadd1}
\sum_{n=1}^{N} \tilde{\gamma}_{n} = 1 \,.
\end{equation}

\noindent
The frequency scale $\gamma_{t}$ defining the dimensionless time $\tau$ 
is just the real part of the factor multiplying $\delta c$ 
after the phase shifts. N-1 of the input channels will have a time-reversed 
counterpart corresponding to the outgoing channels fulfilling
\begin{equation}
\label{eq:gennlinboundary}
\delta c_{out}^{n}  = 
\sqrt{2 \tilde{\gamma}_{n}}\,\delta c - \delta c_{in}^{n}\,.
\end{equation}

\noindent
The remaining input channel will account for the irreversible losses.
The corresponding spectra are related with the intracavity spectra by
\begin{equation}
\label{eq:genspectra}
S_{-,+}^{n}(\tilde{\omega}) = 1 + :S_{-,+}^{n}(\tilde{\omega}):
\; = \; 1 + 2 \tilde{\gamma}_{n} :S_{-,+}(\tilde{\omega}): \,,
\end{equation}
where $:S_{-,+}(\tilde{\omega}):$ denotes the intracavity spectra and 
we have made use of the proportionality of normally ordered 
intracavity and outgoing correlations \cite{Gar91}. The spectra 
$S_{-,+}^{n}(\tilde{\omega})$ coincide with the spectra of the original 
formulation as the new outgoing modes are just a phase sift of the 
originals.

Let us now define a sort of ``reference'' system with only one 
time-reversible input channel, i.e.,
\begin{equation}
\label{eq:refeq}
\frac{d\,\delta c}{d \tau} =   
-\left [1 + i \Delta \right]\delta c + \tilde{B}\,\delta c^{\dagger} 
+ \sqrt{2}\, \delta c_{in}^{ref}\,,
\end{equation}

\noindent
and
\begin{equation}
\label{eq:refbc}
\delta c_{out}^{ref} = \sqrt{2}\, \delta c - \delta c_{in}^{ref} \,.
\end{equation}

\noindent
Obviously $:S_{-,+}^{ref}(\tilde{\omega}): \; = 2 
:S_{-,+}(\tilde{\omega}):$ so that we get finally
\begin{equation}
\label{eq:Scentral}
 S_{-,+}^{n}(\tilde{\omega}) = 1 +
 \tilde{\gamma}_{n} :S_{-,+}^{ref}(\tilde{\omega}): \,.
\end{equation}

\noindent
This is the central result of this section. Let us elaborate a little 
about its interpretation. Squeezing in a given output channel
means that for a certain range of phase shifts the corresponding quadratures
show an intensity of their fluctuations below that of the associated 
incoming channel (assumed in a coherent state). In view of 
Eqs.\ (\ref{eq:gennlinboundary}), the amplitude of the outgoing fluctuations
is a coherent superposition of the intracavity and the incoming 
fluctuations. Squeezing is possible if an adequate correlation between 
$\delta c$ and the relevant input channel is established. But the intracavity 
field is nothing else than the dynamic response of the intracavity system to 
the incoming channels. The input channels are uncorrelated and so the 
dynamic response of the intracavity system to them. A given input 
channel can consequently correlate only with the dynamic response to itself. 
The presence of any other input channel can only degrade the effect.
The great advantage of the tilde normalization is that makes this fact 
explicit. Indeed, Eq.\ (\ref{eq:Scentral}) express the output spectra 
as the dynamic response of the system to an isolated input channel, i.e.,
$:S_{-,+}^{ref}(\tilde{\omega}):$, scaled down by the ``static'' 
contribution to the noise owing to the presence of extra input 
channels. The scale factor $\tilde{\gamma}_{n}$ is just the ratio 
between the coupling constant of the chosen output channel and the 
sum of all of them. 

Eq.\ (\ref{eq:Scentral}) greatly simplifies our task of finding the 
optimum path to maximum noise reduction as we can center our efforts onto the 
simple reference system described by Eqs.\ (\ref{eq:refeq}) and 
(\ref{eq:refbc}). Even more interesting the results concerning the 
reference system will be of general applicability to any one-mode 
system including as such any multiply resonant system under 
adiabatic elimination of all the modes but one. The normally ordered spectra 
of the reference system are easily calculated as
\begin{equation}
\label{eq:SN}
:S^{ref}_{-,+}(\tilde{\omega}):\; = 4 |\tilde{B}| \frac{  
2|\tilde{B}| \mp \sqrt{(1+\tilde{\omega}^{2}+|\tilde{B}|^{2}-\Delta^{2})^{2}
+4 \Delta^{2}}}{(1-\tilde{\omega}^{2}-|\tilde{B}|^{2}+\Delta^{2})^{2}
+4 \tilde{\omega}^{2}} \,.
\end{equation}

\noindent
Our first step is to determine if the dynamic response is capable of 
a total noise suppression. Perfect squeezing can only occur at a dynamic 
instability. Equaling to zero the l.h.s.\ of Eq.\ (\ref{eq:eigen}) 
(the only possible unstable eigenvalue) and after proper normalization an 
equation determining the instability can be written as  

\begin{equation}
\label{eq:refins}
1 + \Delta^{2} = |\tilde{B}|^{2}  \,.
\end{equation}

\noindent
Written in this way an interesting parallelism with the standard OPO 
below threshold shows up, i.e.\, an instability appears when the modulus of 
the ``losses'' coefficient equals that of the ``parametric'' 
coefficient, a sort of natural extension of the condition for the
instability in the conventional OPO for which the coefficients are real. 
Inserting the instability condition (\ref{eq:refins}) in 
$:S^{ref}_{-}(\tilde{\omega}):$ results in
\begin{equation}
\label{eq:Si}
:S^{ref}_{I}(\tilde{\omega}):\; = 4 |\tilde{B}| \frac{  
2|\tilde{B}|-\sqrt{4 |\tilde{B}|^{2} + 
\tilde{\omega}^{2}(\tilde{\omega}^{2}+4)}}{
\tilde{\omega}^{2}(\tilde{\omega}^{2}+4)} \,.
\end{equation}

\noindent
Applying L'Hopital's rule with respect to $\tilde{\omega}^{2}$, 
$:S^{ref}_{I}(\tilde{\omega}):$ equals -1 at 
$\tilde{\omega}=0$, that is, perfect squeezing is obtained at the instability, again in
parallel with OPO. In other words, the dynamic response of the system 
assuming that the condition Eq.\ (\ref{eq:refins}) is reachable, 
is capable of a complete suppression of quantum noise.

Spectrum (\ref{eq:SN}) is simple enough to permit analytical 
optimization. Taking 
partial derivative in $:S^{ref}_{-}(\tilde{\omega}):$ with respect to 
$\tilde{\omega}$ and equaling to zero, $\tilde{\omega}=0$ appears as 
the optimum point whatever the values of $\Delta$ and $|\tilde{B}|$. 
The same applies to $\Delta = 0$ when taking partial derivative with 
respect to $\Delta$. Notice that this last condition implies also a 
squeezing phase independent of the frequency. The optimized noise obtained
imposing these two conditions simplifies to
\begin{equation}
\label{eq:Sopt}
:S_{opt}:\; = -\frac{4 |\tilde{B}|}{(1+|\tilde{B}|)^{2}}\,,
\end{equation}

\noindent
with a minimum at the instability $|\tilde{B}| = 1$ approached 
monotonically. These conditions ($\tilde{\omega}=0$, $\Delta=0$ and 
$|\tilde{B}| = 1$) will help us in finding optimum paths. In 
particular, moving $|\tilde{B}|$ from zero to one while maintaining 
$\Delta=\tilde{\omega}=0$ defines an optimum path reaching the 
instability for the ``reference'' model.

An optimum path is defined solely by the squeezing spectrum leaving 
aside the ``stretching'' one. It is important to study also the accompanying 
excess noise on the conjugate quadrature for it could invalidate in 
practice the optimum path if this excess noise is unbearable high. 
The minimal excess noise production imposed by the Heisenberg
principle corresponds to $S_{-}(\omega)S_{+}(\omega)=1$. It is a perfect
complementary relation between quadratures: the deamplification of
fluctuations in a given quadrature must equal the amplification of 
fluctuations in the conjugate. In that case we are dealing with a Minimum 
Uncertainty State (MUS) for those quadratures, the text-book 
definition of a squeezed state. Adding 1 to Eq.\ (\ref{eq:SN}) and 
after some minor algebra
\begin{equation}
\label{eq:SS}
S_{-,+}^{ref}(\tilde{\omega}) = 
\frac{\left (
2 |\tilde{B}| \pm \sqrt{(\tilde{\omega}^{2}+|\tilde{B}|^{2}+1-\Delta^{2})^{2}
+4 \Delta^{2}} 
\right )^{2}}
{(1-\tilde{\omega}^{2}-|\tilde{B}|^{2}+\Delta^{2})^{2}
+4 \tilde{\omega}^{2}} \,.
\end{equation}

\noindent
Straightforward algebra leads to $S_{-}^{ref}(\tilde{\omega}) 
S_{+}^{ref}(\tilde{\omega}) =1$. The excess noise is minimum again 
in parallel to the standard OPO system with real coefficients. 

\subsection{Standard normalization}
As discussed in the introduction we are here principally interested in 
the squeezing behavior with respect to the photon number $n$.
Unfortunately the tilde normalization is inappropriate to such a task 
as the frequency scale depends on $n$ itself.
It is far more convenient to use $\gamma^{-1}$ as the time scale instead of
$\gamma_{t}^{-1}$ and to normalize the photon number as $m = \nu
n/\gamma$. In complete parallelism to the tilde normalization we have
then,
\begin{eqnarray}
\label{eq:normeqhat}
\frac{d\,\delta c}{d \tau} & = &  
-\left [1 + 2 m K_{r} + i (\hat{\delta} + 2 m K_{i}) \right]\delta c 
+ \left [\sqrt{K_{r}}\,\eta_{in} - (K_{r}+i K_{i}) m \right ]\,
\delta c^{\dagger} \nonumber \\
& + & 2 \sqrt{m K_{r}}\;\delta d_{in} +
\sqrt{2 \hat{\gamma}_{c}}\;\delta c_{in} +
\sqrt{2\hat{\gamma}_{s}}\,s_{in}\,,
\end{eqnarray}
where the hat represents divided by $\gamma$ and
\begin{equation}
\label{eq:etain}
\eta_{in} \equiv \frac{2 \sqrt{\nu}}{\gamma}\,\beta_{in}e^{-i2\theta}\,,
\end{equation}

\noindent
which represents the harmonic mode input amplitude normalized to the 
value at the standard OPO threshold. The spectra (\ref{eq:spectrum}) 
become now
\begin{mathletters}
\label{eq:hatspectrum}
\begin{eqnarray}
S^{a}_{-,+}(\hat{\omega}) & = & 1 + 4 \hat{\gamma}_{c}
 |\hat{B}| \frac{\hat{N}_{-,+}}{\hat{D}}\,,
\label{eq:hatspectruma} \\ 
S^{b}_{-,+}(\hat{\omega}) & = & 
1 + 8 K_{r}m |\hat{B}| \frac{\hat{N}_{-,+}}{\hat{D}}\,,
\label{eq:hatspectrumb} 
\end{eqnarray}
\end{mathletters}

\noindent
with
\begin{mathletters}
\label{eq:hatSc}
\begin{eqnarray}
\hat{N}_{-,+} & = & 2 |\hat{B}| (1 + 2 K_{r} m) \mp 
\sqrt{\left [ (1 + 2 K_{r} m)^{2} - (\hat{\delta} + 
2 K_{i} m)^{2}+ |\hat{B}|^{2} +\hat{\omega}^{2} \right ]^{2} + 
4 (1 + 2 K_{r} m)^{2}
(\hat{\delta} +2 K_{i} m)^{2}}\,,
\label{eq:hatScN}\\
\hat{D} & = &  \left [(1+2 K_{r} m)^{2}+(\hat{\delta}+ 2 K_{i} m)^{2} 
- |\hat{B}|^{2} - \hat{\omega}^{2} \right ]^{2}+ 
4 (1+2 K_{r} m)^{2} \hat{\omega}^{2} \,,
\label{eq:hatScD}
\end{eqnarray}
\end{mathletters}

\noindent
and $\hat{B} = \sqrt{K_{r}}\,\eta_{in} - (K_{r}+i K_{i}) m$.
We shall refer to the above formulae 
as the hat normalization. To refer the frequency to the cavity decay 
constant with the subsequent re-normalization of the system 
parameters, i.e.\, the hat normalization, is quite a standard procedure 
in the literature.

Some care must be taken when studying the quantum noise behavior as a 
function of $n$ (or $m$). It is not a free parameter of the problem 
as would be the input fields or the phase mismatch but it is in a nonlinear
relation with them. We need, therefore, to check that the proposed 
values of $n$ are indeed a solution of Eq.\ (\ref{eq:neqn}).
Fortunately the spectra (\ref{eq:spectrum}) do not depend on the overall 
sign of $\delta + 2 \Gamma n$ and therefore the conclusions reached in section 
\ref{LEE} about the existence of $|\alpha_{in}|$ permit a safe 
variation of $n$ in search of strong noise reduction provided the 
stability of the corresponding fixed points.

\subsection{Squeezing at the fundamental mode}
Applying Eq.\ (\ref{eq:Scentral}) to the fundamental mode
\begin{equation}
\label{eq:Satilde}
S^{a}_{-,+}(\tilde{\omega}) = 1 + \tilde{\gamma}_{c} 
:S^{ref}_{-,+}(\tilde{\omega}): \,.
\end{equation}

\begin{figure}
\centerline{\includegraphics*[scale=0.5]{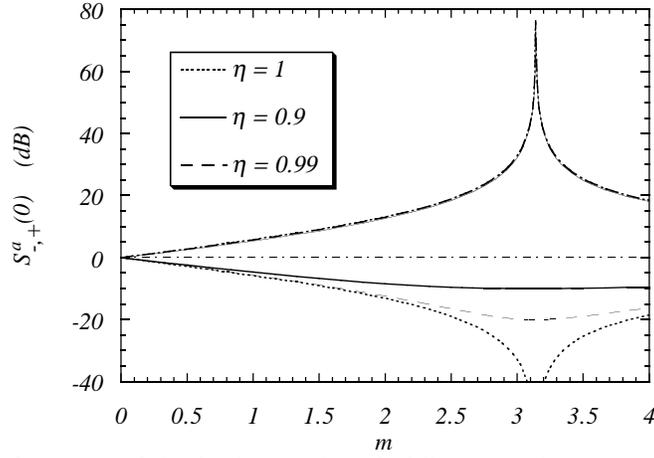}}
\caption{Noise spectra at zero frequency of the fundamental mode following and optimum 
path for three escape efficiencies of the cavity including the ideal 
case $\eta = 1$. The curves above the 
divergences are not physical.}
\label{fg:SaKerr}
\end{figure}

\noindent
It is clear that the best
performance corresponds to $\gamma_{nl} = 0$, that is, either $n=0$ or $\mu = 
0$, as then $\tilde{\gamma}_{c}$ maximizes to 
$\eta \equiv \gamma_{c}/(\gamma_{c}+\gamma_{s})$
(the escape efficiency of the cavity). The case $n=0$ corresponds to 
the very well known case of squeezed vacuum generation. For $\mu =0$ 
and finite $n$ the system is formally equivalent to a resonant optical 
Kerr effect system whose quantum noise behavior has been amply studied 
previously \cite{Rey89}. The condition (\ref{eq:refins}) reduces for $\mu =0$
to $\delta = -2 n \Gamma \pm \sqrt{n^{2}\Gamma^{2}-\gamma^{2}}$, the 
well known turning points of optical dispersive bistability \cite{Rey89} 
but with the nonlinear dispersion induced by cascading. Indeed, such cascading 
induced bistability has been experimentally demonstrated in \cite{Whi96a}. 
Rewriting it within the hat normalization the condition reads 
\begin{equation}
\label{eq:insKerr}
\hat{\delta}_{\pm} = -2 m K_{i} \pm \sqrt{m^{2} K_{i}^{2} - 1} \,,
\end{equation}

\noindent
where $K_{i}=-1/\pi$ as it is evaluated at $\Delta k L_{m} = 2 \pi$.
Once $\tilde{\gamma}_{c}$ is independent of $\Delta k$ and $n$, 
the optimum path with respect to $m$ (the only remaining free parameter)
is determined solely by the reference system. It corresponds to increase 
$m$ till $m=\pi$ (where the condition (\ref{eq:insKerr}) is reached) 
while maintaining $\hat{\delta} = 2 m / \pi$ and $\omega = 0$. 
Fig.\ \ref{fg:SaKerr} displays the evolution of 
both the maximum squeezing and the maximum excess noise following 
such a path for three values of the escape 
efficiency, namely, 0.9, 0.99 and the ideal 1. The noise is expressed in 
dB's with respect to the vacuum noise. A Heisenberg limited excess noise 
appears in such a case as a specular image of the squeezing. The 
instability is signaled by the divergence in the excess noise. 
Above it, the curves shown are not physical as they correspond to 
unstable fixed points.
The case $\eta = 0.99$ in Fig.\ \ref{fg:SaKerr} shows an excellent 
behavior with an almost Heisenberg limited excess noise till 
near the instability.

\begin{figure}
\centerline{\includegraphics*[scale=0.5]{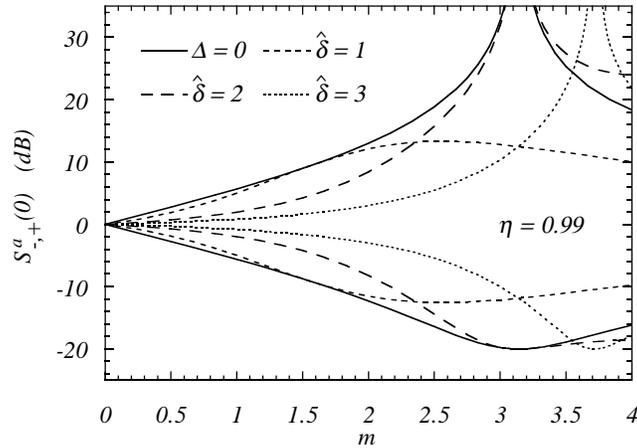}}
\caption{Comparison among noise spectra at zero frequency (fundamental 
mode) following various paths in the parameter space. 
The optimum one corresponds to $\Delta = 0$. Above the divergences 
the results are not physical.}
\label{fg:SaKcomp}
\end{figure}

Fig.\ \ref{fg:SaKcomp} illustrates the idea of optimum path by comparing 
the $\eta=0.99$ plot of Fig.\ \ref{fg:SaKerr} against various cases 
with fixed values of $\hat{\delta}$. Below $m=\pi$, for a given $m$ 
the maximum squeezing is obtained when $\Delta = 0$ as expected. 
Above $m=\pi$ it is not possible to reach the minimum noise of 
$1-\eta$ fulfilling $\Delta = 0$.  

Being $\mu =0$ the formulae simplify enough for allowing a simple 
expression for the squeezing phase. More specifically, from 
$B=-i\Gamma \alpha^{2}$, $\theta_{m}=\theta + \pi/2$. On the other 
hand substituting Eq.\ (\ref{eq:state}) in Eq.\ (\ref{eq:boundarya}) results in  
$\sqrt{2 \gamma_{c}}\alpha_{out} = \alpha (\gamma_{c}-\gamma_{s} + i \Gamma 
n)$, giving a squeezing phase relative to that of the output field of
\[
\frac{\pi}{2} - \arctan \left (\frac{\Gamma n}{\gamma_{c}-\gamma_{s}} \right 
)\,.  
\]

\noindent
At the instability $\Gamma n = \gamma$ and for low 
$\gamma_{s}$ it approaches $45^{o}$. There is a possibility of an extra
control of the squeezing phase not present in the conventional Kerr 
effect system by making use of the harmonic mode. Taking a finite 
$\mu$ but low enough so that $\mu n \ll \gamma$ and at the same 
time a $\beta_{in}$ high enough to imply $2\sqrt{\mu} \beta_{in} \approx 
\gamma$, still we will have $\gamma_{nl} \approx \gamma$ while the 
squeezing phase relative to the output field will depend on both the 
modulus and the relative phase between the input fields. In practice, however, 
maintaining $\mu n$ very low could imply a exceedingly high $\beta_{in}$ 
in order to have a $2\sqrt{\mu}\beta_{in}$ intense
enough for a significant influence on the final phases. Of course, the 
instability point would accordingly depend on $\beta_{in}$.
      
\begin{figure}
\centerline{\includegraphics*[scale=0.5]{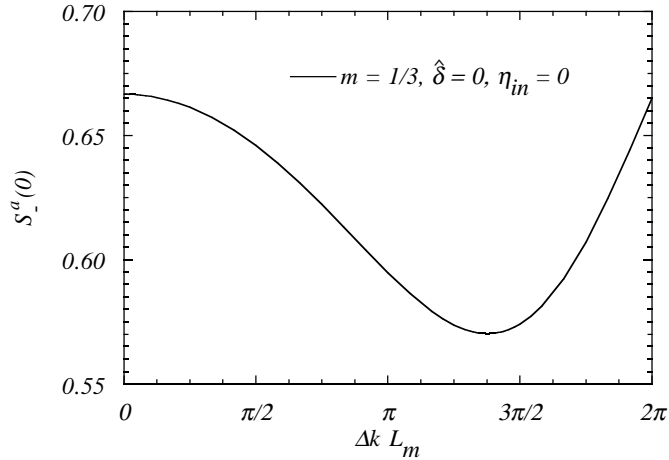}}
\caption{Squeezing in the fundamental mode at zero frequency as a 
function of the phase mismatch for a low intracavity photon number.}
\label{fg:Sa}
\end{figure}

There is no hope of any behavior similar to the reported in \cite{Cab97}
as a competition between second and third order nonlinearities needs
$\mu \neq 0$, opening the fundamental mode to the fluctuations of the
input harmonic mode with strong deleterious effects. At most, some
remnants of the
enhanced efficiency coming from the competition between nonlinearities can be 
observed for low $\gamma_{nl}$. Then, as shown in 
Fig.\ \ref{fg:Sa}, the best working point is not necessarily located at
$\mu = 0$, i.e.\, maximum squeezing is obtained with a finite mismatch.

\subsection{Squeezing at the harmonic mode} 
For the harmonic mode Eq.\ (\ref{eq:Scentral}) yields
\begin{equation}
\label{eq:Sbtilde}
S^{b}_{-,+}(\tilde{\omega}) = 1 + \tilde{\gamma}_{nl} 
:S^{ref}_{-,+}(\tilde{\omega}): \,.
\end{equation}

\noindent
Now, the situation is the complete opposite: the performance is favored
by a finite $\mu$ in order to have a non-zero $\gamma_{nl}$ and a large
$n$ to approach the ratio $\tilde{\gamma}_{nl}$ to one. In fact, under
ideal conditions of perfect dynamic noise suppression and no absorption
and scattering losses ($\gamma_{s}=0$), the squeezing in both modes 
are complementary in the sense of 
\begin{equation}
\label{eq:abcomple}
S^{a}_{-}+S^{b}_{-} = 2
-\frac{\gamma_{c}}{\gamma_{t}}-\frac{\gamma_{nl}}{\gamma_{t}} \;=\; 1\,,
\end{equation}

\noindent
a direct consequence of the fluctuation-dissipation relation 
(\ref{eq:genadd1}). This complementarity has been previously reported for
the doubly resonant degenerate parametric oscillator \cite{Fab90}.
The maximum squeezing available for the harmonic mode whatever the 
dynamic response of the system is easily obtained by setting
$:S^{ref}_{-,+}(\tilde{\omega}):$ to -1 in Eq.\ (\ref{eq:Sbtilde}), that is,
\begin{equation}
\label{eq:ilimit}
S_{M} = 1 - \frac{2 m K_{r}}{1+2 m K_{r}} \; = \; 
\frac{1}{1+2 m K_{r}} \,.
\end{equation} 

\noindent
This static contribution to the noise is now nonlinear in the sense 
that it depends on the phase mismatch and $m$. An immediate consequence 
of Eq.\ (\ref{eq:ilimit}) is the possibility of an arbitrarily large 
quantum noise reduction for any finite value of $K_{r}$. The 1/9 limit 
of the conventional phase-matched SHG is therefore due to a failure 
of the setup to maximize the dynamic response of the system. Let us 
center then, firstly in the SHG-like case with $\beta_{in}=0$ as it 
includes the above mentioned conventional setup (the experiments in 
\cite{Pas94} and \cite{Tsu95}). The instability
points are now given by (directly in the hat normalization)
\begin{equation}
\label{eq:hatinsb}
\hat{\delta}_{\pm} = - 2 m K_{i} \pm \sqrt{m^{2} 
(K_{i}^{2}-3 K_{r}^{2}) - 4 K_{r} m - 1} \,.
\end{equation}

\noindent
Both kinds of nonlinearities (dispersive and absorptive) are in this case 
necessary as the factor $K_{i}^{2}-3 K_{r}^{2}$ needs to be positive to 
allow $\hat{\delta}_{\pm}$ to be real. The phase-matched case is 
therefore excluded.
\begin{figure}
\centerline{\includegraphics*[scale=0.5]{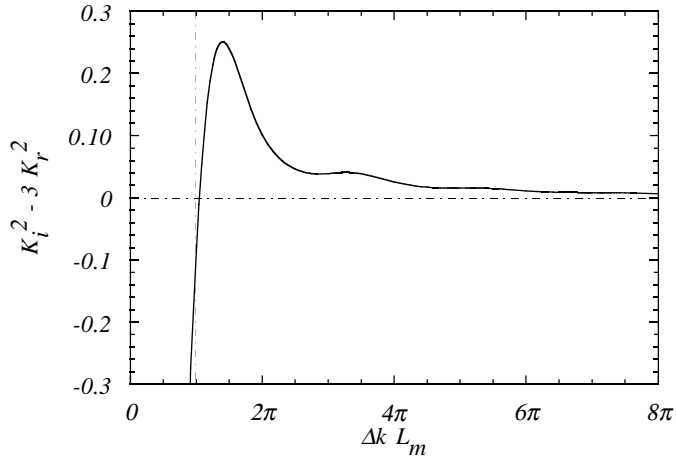}}
\caption{The value of $K_{i}-3 K_{r}$ as a function of the phase 
mismatch.}
\label{fg:g-3f}
\end{figure}

\begin{figure}
\centerline{\includegraphics*[scale=0.5]{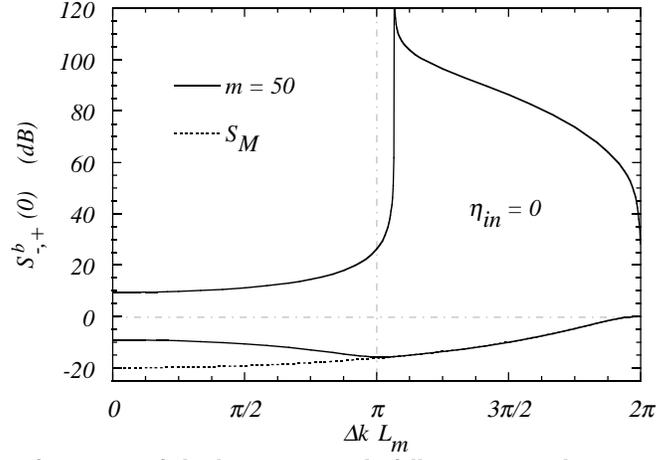}}
\caption{Noise spectra at zero frequency of the harmonic mode 
following a nearly optimum path with respect to the phase mismatch for the 
SHG like case.}
\label{fg:Sbkm50}
\end{figure}

\noindent
Fig.\ \ref{fg:g-3f} shows $K_{i}^{2}-3 K_{r}^{2}$ as a function of the
phase mismatch and indeed near above $\pi$ it is positive. Optimum 
approaches to (\ref{eq:hatinsb}) are now more difficult to 
evaluate than in the fundamental mode as both the dynamic processing of 
the noise and the static contribution from the noise inputs (encoded 
in $\tilde{\gamma}_{nl}$) depend on $m$ and $\Delta k$. With respect to 
to $m$ is clear that the static part is optimized at $m \rightarrow 
\infty$. This limit can be approached letting $\tilde{\omega} = 0$ 
and $\hat{\delta} = - 2 K_{i} m$ (i.e.\, $\Delta = 0$). $|\tilde{B}|$ 
reduces in this case of $\beta_{in}=0$ to $m 
\sqrt{K_{r}^{2}+K_{i}^{2}}/(1+2 m K_{r})$ showing a monotonic 
increasing behavior with respect to $m$ from 0 to the maximum
(at $m \rightarrow \infty$)
\begin{equation}
\label{eq:Bmax}
|\tilde{B}| = \frac{1}{4}\left [1+\left (\frac{K_{i}}{K_{r}}\right )^{2} 
\right ] \,.
\end{equation}
Notice that for $K_{i}^{2}=3 K_{r}^{2}$ it consistently equals 1. 
We have then, both $\tilde{\gamma}_{nl}=1$  and the fastest approach to 1 of 
$|\tilde{B}|$ when $m \rightarrow \infty$. Therefore, the squeezing 
along an optimum path with respect to $\Delta k$ is given by substituting 
Eq.\ (\ref{eq:Bmax}) in the spectrum (\ref{eq:Sopt}) and then the 
obtained $:S_{opt}:$ in Eq.\ (\ref{eq:Sbtilde}). Obviously, in a 
real experiment $m$ can be large but always finite. Let us take as a 
``large'' $m$ one giving a $S_{M}$ around 20 dB as in the 
case $\eta = 0.99$ of Fig.\ \ref{fg:SaKerr}. This corresponds to 
$m=50$. Figure \ref{fg:Sbkm50} displays $S^{b}_{-,+}$ as a function 
of the phase mismatch in such a case.
To illustrate the modulation exerted by $S_{M}$ we have take this time 
$\hat{\delta}$ equal to the real part of Eq.\ (\ref{eq:hatinsb}) plus a 
very small number. In this way the plot remains valid for the whole range 
of the $\Delta k L_{m}$. While Eq.\ (\ref{eq:hatinsb}) is complex the condition 
$\Delta =0$ is almost fulfilled and above the instability the noise 
suppression reduction follows $S_{M}$. Again the pernicious effect of 
the instability regarding the excess noise has a very short range. 
For comparison $S_{M}$ is also depicted.

\begin{figure}
\centerline{\includegraphics*[scale=0.5]{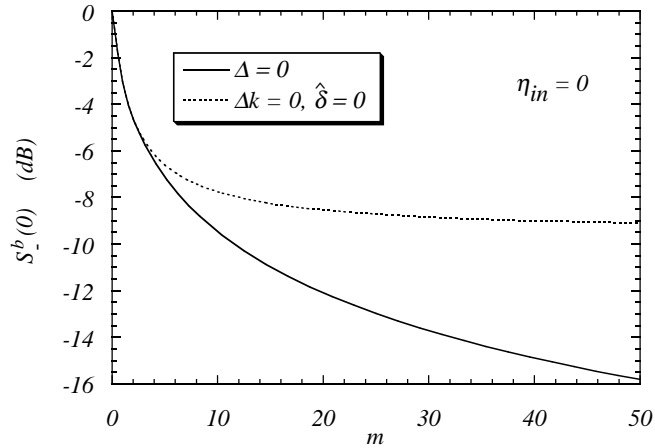}}
\caption{Squeezing in the harmonic mode along an optimum path with 
respect to the normalized intracavity photon number ($m$) for the SHG 
like case compared with the phase-matched SHG case.}
\label{fg:Sboptbin0}
\end{figure}

The optimum path with respect to $m$ is much more complicated to find 
because the intricate dependence of $K_{i}$ and $K_{r}$ with respect 
to the phase mismatch. Figure \ref{fg:Sboptbin0} has been generated 
finding numerically the minima of $S_{-}^{b}(0)$ while scanning 
the range of $m$. For comparison the phase-matched case is also depicted 
showing an asymptotic behavior towards $-10 \log 9$. For low values 
of $m$ the effect of $\tilde{\gamma}_{nl}$ overwhelms the dynamic 
response so that the best value corresponds to maximize $K_{r}$. 
As soon as the two curves depart from each other the dynamic response 
dominates the behavior and the minimum noise is at the instability as in 
Fig.\ \ref{fg:Sbkm50}. At this stage the optimum path begins to follow 
the instability all the time. It should be taken then, as a 
mathematical limit. However, in view of Fig. \ref{fg:Sbkm50}, before 
reaching it, bearable values of the excess noise are accessible with 
a slight diminution of the squeezing.

It is worth to mention that a squeezing as large as 
48\% induced by cascading has been very recently reported \cite{Kas97}.
The cascading was due, however, to a detuning of the pump mode in 
a triply resonant non-degenerate OPO with a much lower finesse for the 
pump mode rather than by phase mismatch. 
Under such conditions a cascaded $\chi^{(3)}$ is also 
induced leading ideally to perfect squeezing in the pump mode.  

Although a finite mismatch allows to reach $S_{M}$,
the overall optimum working point corresponding to 
$\Delta k = 0$ is out of reach. The question arises then, of if it is
possible to fulfill Eq.\ (\ref{eq:refins}) at $\Delta k = 0$. A glance 
at the definition of $B$ suggests it should be possible adding a 
driving to the harmonic mode. In such a case $\tilde{\gamma}_{nl}$ simplifies 
to $2m/(1+2m)$ obviously independent of $\eta_{in}$. Constructing an optimum path 
with respect to the harmonic input reduces then, to set $\Delta=\tilde{\omega}=0$. 
Phase matching along with $\Delta = 0$ implies $K_{r} = 1$, $K_{i} = 0$ and 
$\hat{\delta} = 0$ so that the instability condition (\ref{eq:refins}) 
simplifies to
\begin{equation}
\label{eq:etains}
1 + 2 m = |\eta_{in} - m| \,,
\end{equation}

\noindent
a perfectly achievable condition. 
We can further optimize by choosing the phase of $\eta_{in}$ 
adequately to approach $\tilde{B}$ to one as much as possible.
The extreme cases correspond to $\eta_{in}$ real, i.e., 
$\eta_{in}= - (1 + m)$ and $\eta_{in}= 1 + 3 m$.   
From Eqs.\ (\ref{eq:phases}) and (\ref{eq:ain}) it is easy to check 
that they correspond respectively to
$\phi-\varphi/2 = \pi$ and $\phi-\varphi/2=0$. The negative case 
maximizes $\tilde{B}$. It has been previously reported in \cite{Sch95}. 
Taking squared modulus of Eq.\ (\ref{eq:boundaryb}) 
the negative case appears as promoting harmonic output power 
while the converse is valid for the positive. 
The squeezing phase is also easy to calculate in this case. 
In particular, given the correlation (\ref{eq:corrb}), $\nu(\omega)$ 
is determined by the phase of $\alpha^{2} B$ (independent of $\omega$ 
as $\Delta =0$), something proportional to
\[
(\eta_{in}-m) e^{i 4 \theta} \,.
\]

\noindent
The corresponding squeezing phases are $\theta_{m} = 2\theta + \pi$ 
for the negative case while for the positive case it changes from  
$\theta_{m} = 2\theta + \pi$ to $\theta_{m} = 2\theta + \pi/2$ at 
$\eta_{in}=m$. On the other hand, the output harmonic amplitude is 
proportional to (see Eq.\ (\ref{eq:boundaryb}))
\begin{equation}
\label{eq:bout}
b_{out} \propto (\eta_{in} - 2 m) e^{i (2 \theta + \pi)} \,.
\end{equation}

\begin{figure}
\centerline{\includegraphics*[scale=0.5]{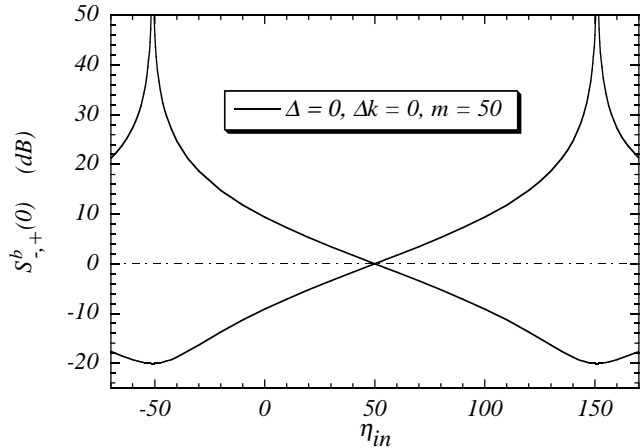}}
\caption{Noise spectra at zero frequency (harmonic mode) following an 
optimum path with respect to the normalized input harmonic amplitude
($\eta_{in}$). The curves are not physical 
above the divergences.}
\label{fg:Sbin50}
\end{figure}

\noindent
Consequently, the relative squeezing phase for the negative 
$\eta_{in}$ is $\pi$, i.e., amplitude squeezing. The positive case is 
more complicated. It remains equal $\pi$ (amplitude squeezing) till 
$\eta=m$. Above this value it changes to $\pm \pi/2$ depending on the 
sign of $\eta_{in}/2 - m$ yielding in any case phase squeezing. At a 
first glance, it appears there is a sudden change from amplitude to 
phase squeezing when the input phases are fix to $\phi-\varphi/2=0$ 
and $|\eta_{in}|$ passes through $m$. It is not so however, as at 
this point $B=0$ and the state collapses to a coherent state with no 
squeezing. The situation is clearly depicted in figure \ref{fg:Sbin50} 
where $S_{-,+}^{b}(0)$ are displayed as a function of $\eta_{in}$ 
assumed real. The r.h.s.\ of the plot corresponds to $\phi-\varphi/2 = 0$
while the l.h.s.\ to $\phi-\varphi/2 = \pi$ and for negative ordinates 
it should be considered as an optimum path with respect to $\eta_{in}$. 
The behavior is completely symmetric with respect to $\eta_{in}=m$ 
where both the squeezing and the excess noise equal that of the vacuum.

\begin{figure}
\centerline{\includegraphics*[scale=0.5]{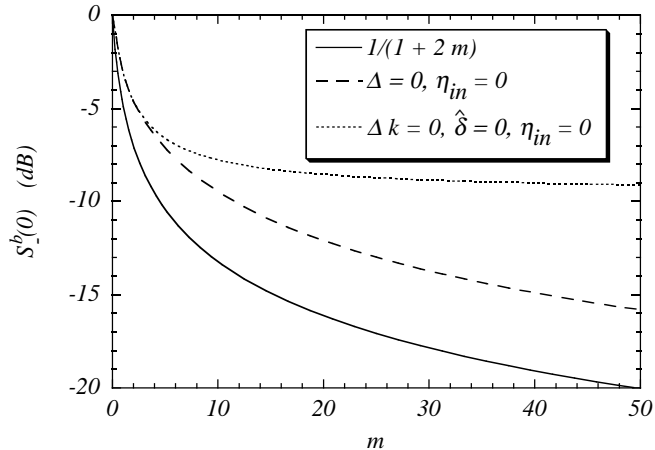}}
\caption{Maximum squeezing (harmonic mode) as a function of the 
normalized intracavity photon number ($m$) in nonlinear second order singly 
resonant device. For comparison the phase-matched and optimized SHG 
cases are also shown.}
\label{fg:SM}
\end{figure}

The optimum path with respect to $m$ is now given by 
$S_{M}$ at $K_{r}=1$. Figure \ref{fg:SM} is the equivalent to 
Fig.\ \ref{fg:Sboptbin0} for the new situation. It represents the 
maximum efficiency as far as quantum noise reduction is concerned the 
system can yield in any way with respect to $m$. The 
improvement with respect to the standard phase-matched SHG as well as 
to the optimized SHG is certainly high.

\section{Discussion and conclusions}{\label{DC}}
Two are the main purposes of the present work. On one side, to gain 
physical insight about the origins of quantum noise in singly 
resonant systems. On the other, to explore their potential as 
squeezed light sources. In such a task we have used a model including 
all the relevant physics we wanted to address but simple enough to 
be tractable. The results shown in the previous sections 
certainly reveal a high potential of the studied configurations. 
An evaluation of the limits of the model in reproducing the real 
physical situation as well as a discussion of possible implementations, 
seems, therefore, in order.

One obvious idealization of the model is to assume perfectly coherent 
inputs neglecting the excess noise of real lasers something
expected to have deleterious effects at low frequencies.  White and 
coworkers \cite{Whi96b} have developed an analytical approach to
this problem resulting in an impressive agreement with the experiments. 
As expected, the excess noise completely destroys the 
squeezing at low frequencies. In their experiments, however,
the deleterious effect was restricted to only 7 MHz by adding
a mode cleaner to the system, the spectrum coinciding with the ideal 
one out of this range. Even better, in \cite{Bre97} the laser noise 
was shot-noise-limited  down to 1 MHz, again using an external mode cleaner.

Considering as sensible the assumption of coherent states for the input 
modes as well as a value of $m$ around 3 (we will see below it looks 
like the case) our main concern about the fundamental mode 
results summarized in Fig.\ \ref{fg:SaKerr} refers to the feasibility 
of the chosen escape efficiencies.  The ratio 
$\gamma_{c}/(\gamma_{c}+\gamma_{s})$ is difficult to maximize in a resonant 
mode because, by its own resonant nature, $\gamma_{c}$ must be rather low. 
Thus, in \cite{Whi96a} it was
only of 0.52, while in \cite{Pas94} it was 0.36. Even in \cite{Kur93}, a
doubly resonant system specifically designed to squeeze the fundamental
mode, the escape efficiency was around 0.9, limiting the maximum
squeezing achievable to 90\% (in practice, a 52\% of noise reduction was
reached). It appears, then, that nowadays the $\eta=0.99$ should be taken 
rather as an ideal illustrative case.

In contrast, the ultimate limit for the noise 
suppression in the harmonic mode (Eq.\ (\ref{eq:ilimit})) is pushed up 
by the fundamental mode photon number, opening a way to bypass the 
usual untouchable limit imposed by the escape efficiency of the cavity
(as in the fundamental mode). Therefore, the squeezing in the harmonic mode 
can be arbitrarily large under the ideal assumption that the energy 
load inside the cavity can be also arbitrarily large. However, this is 
not totally true as the model does not take into 
account the losses in the harmonic mode which necessarily limit the 
degree of noise suppression. We can estimate this 
limitation assuming the absorption in one single pass through the 
nonlinear material equivalent to the effect of a beam splitter with 
the adequate reflectivity. Taking an absorption of 0.6\%/cm as in 
\cite{Kur93} and a length of 1 cm, the equivalent reflectivity would be 
of $6 \;10^{-3}$. The spectrum after the beam splitter is given by 
$S_{out}=1 + T :S_{in}:$. Setting $:S_{in}: = -1$ and $T = 1-R$, 
the ultimate squeezing achievable is precisely $R = 6\;10^{-3}$, i.e.\, 
- 22 dB. In other words, the chosen value of $m=50$ in 
Figs.\ \ref{fg:Sbkm50} and \ref{fg:Sbin50} represents 
more or less the maximum the model can stand without the inclusion 
of the harmonic mode losses.

Of course, we still cannot assume $m=50$ as a realistic limit for 
the state of the art devices as $m$ depends not only on the 
intra-cavity photon number but on the ratio $\nu/\gamma$ 
between the nonlinearity and losses. This ratio must be high enough in 
order to prevent a degradation of the nonlinear optical response in 
the system as commented in the introduction. 
Besides, this ratio scales down the power 
available in the external sources. In view of these complications, probably 
the most reliable way of setting the physical scale of $m$ is to 
compare the results with the reported experiments. In \cite{Tsu95} 
the quoted noise reduction was of -5.2 dB. Setting to zero 
$\Delta k$ and $\beta_{in}$ in Eq.\ (\ref{eq:hatspectrumb}) corresponding 
to phase-matched SHG, a -5.2 dB squeezing results at $m=2.5$, far from 
the $m=50$ limit. Fortunately, the limit 
(\ref{eq:ilimit}) grows up quite quickly for low $m's$ 
(see Fig.\ \ref{fg:SM}). Thus, 10 dB of noise suppression are 
reached at $m = 5$, no such unthinkable value.
However, -15 dB of noise reduction requires $m=15$, while a 
-20 dB figure is at the $m=50$ limit, an order of magnitude higher. 
New nonlinear materials seem the only possibility for such high 
squeezing degrees. A promising via consists in the use 
of resonant nonlinearities in asymmetric quantum wells (AQW). 
Huge nonlinearities have been demonstrated in frequency 
doubling experiments and even a tuning of the nonlinearity with a d.c. 
field \cite{Sir92}. Obviously, also the absorption is enhanced by the 
resonance. This can be a problem as the ratio $\nu/\gamma$ could at 
the end of the day not be increased. To asses this possibility requires 
quite a detailed analysis out of the scope of the present work.
We can foreseen, however, a promising advantage in the fact that the 
losses in the harmonic mode have little influence on the 
performance. By maintaining a strong two photon resonance but 
relaxing the one photon counterpart (tuning 
with a d.c. field or by an adequate energy level engineering), the 
nonlinearity would be certainly enhanced while the losses at the 
fundamental mode would not increase so strongly, thus enhancing 
$\nu/\gamma$. With only one passage through the cavity of the harmonic
mode and taking into account that a very thin layer of material is capable of SHG 
\cite{Sir92}, the corresponding deleterious effect cannot be very large.
Even a more exciting possibility comes from the recent experimental 
demonstrations of absorption inhibition in AQW induced by quantum 
interference \cite{Fai96,Schm97,Fai97}. The absorption 
transparency and the resonant enhancement can be combined using an 
adequate quantum well engineering leading to very efficient frequency 
doublers (see \cite{Schm96}, where precisely a scheme only resonant at
the harmonic mode is proposed).

These are certainly promising perspectives but we should not dismiss the 
improvements arising at the range of the present nonlinear crystals 
performances. Let us center then, around $m=2.5$.
As shown in the previous section the best strategy corresponds to drive 
both modes with relative phases $\varphi-\phi/2 = \pi/2$ (negative 
$\eta_{in}$) and $\Delta=0$.
\begin{figure}
\centerline{\includegraphics*[scale=0.5]{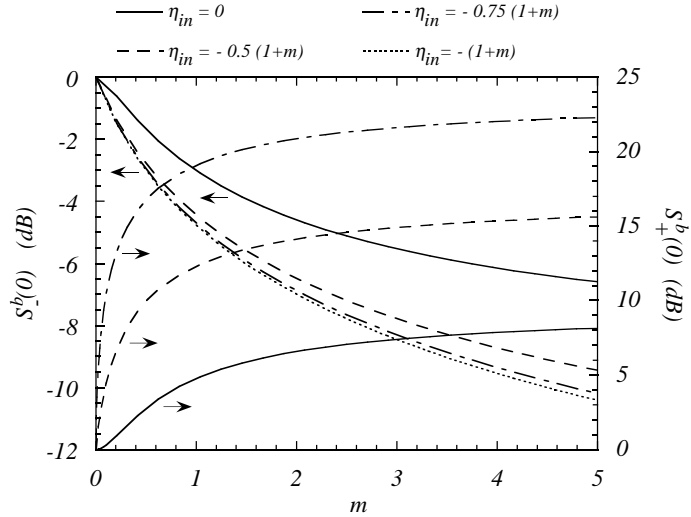}}
\caption{Noise spectra at zero frequency (harmonic mode) for $\Delta 
k = \hat{\delta} = 0$ as function of the normalized intracavity photon 
number at various ``distances'' from the dynamic instability.}
\label{fg:sqm5}
\end{figure}

\begin{figure}
\centerline{\includegraphics*[scale=0.5]{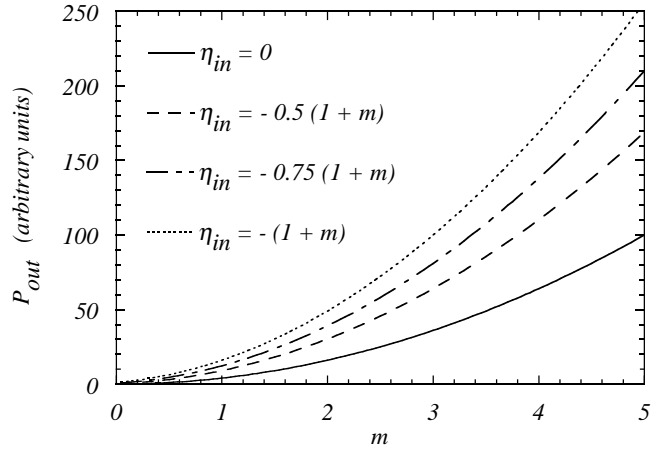}}
\caption{The harmonic output power corresponding to the cases of 
Fig.\ \ref{fg:sqm5}.}
\label{fg:Pw}
\end{figure}

\noindent 
In Fig.\ \ref{fg:sqm5} the noise behavior till $m=5$ is displayed for 
various ``distances'' to the instability (\ref{eq:etains}). 
Even at half the instability $\eta_{in}$ value, 
the squeezing at $m=2.5$ grows from -5.1 dB (69\%) 
to -7.2 dB (80\%). The excess noise, on the other hand, rapidly 
increases at low $m's$ but it also saturates quickly to bearable values.
The improvement, although nothing spectacular is quite substantial. 
In \cite{Sch95} it was not reckoned so because the noise suppression was 
studied as function of the input power. Given its nonlinear relation 
with $m$ the improvement is much slower with respect to this variable. 
Besides the squeezing, the output power is also enhanced. Taking a 
negative $\eta_{in}$ in Eq.\ (\ref{eq:bout}), the output power results in
\begin{equation}
\label{eq:Pw}
P_{out} \propto (2 m + |\eta_{in}|)^{2} \,,
\end{equation}

\noindent
and thus, the harmonic mode input contributes constructively to it.
As shown in Fig.\ \ref{fg:Pw} at half a way of the instability the 
power is nearly doubled. Although from the theoretical point of view 
the injection of a coherent signal in the harmonic mode looks quite 
harmless, the experimental implementation is not trivial. However, the 
remarkable achievements in \cite{Sch96a,Bre97} with the OPA strongly 
support the feasibility of the idea.   

Finally, a word of caution about the design of the device.
It is important to avoid the setting of oscillations out of 
the fundamental mode (the so called subharmonic pumped 
OPO \cite{Sch96b,Sch97}), something capable of 
destroying the noise reduction \cite{Whi97}. At a first glance,
finite values of $\eta_{in}$ would favor the effect by promoting
the down conversion. But it is not necessarily so as the down
conversion is encouraged only for a given range of the relative phase 
between the two driving fields. Thus, for the negative $\eta_{in}$ 
case studied above, being the harmonic output power maximized, the  
down conversion is minimized.

To conclude, let us summarize the most relevant results. 
Firstly, for any system with only 
one effective mode we have given a systematic approach capable of 
isolating the processing of quantum noise by the dynamic response of 
the system. This dynamic processing is maximized 
at zero frequency, zero generalized nonlinear detuning ($\Delta$ as 
defined in section \ref{SP}) and at a dynamic instability. 
The static contributions to the noise coming from the 
different noisy inputs can in some cases, move the overall optimum 
working point away from that corresponding to maximum dynamic noise 
suppression. In spite of such, to have a rule to maximize the dynamic 
quantum noise suppression resulted very useful to characterize the 
squeezing behavior when applied to a specific optical system. 
In particular, for the case of a singly resonant second 
order nonlinear device, the squeezing 
at the fundamental mode is limited by the escape efficiency of the 
cavity, the best working point being within figures of merit 
of conventional nonlinear crystals. In the harmonic 
mode high squeezing requires new materials but it is only limited 
by the losses in the non-resonant harmonic mode opening the possibility 
of using multiple quantum wells with resonantly enhanced nonlinearities. 
However, with standard nonlinear crystals still is possible a 
substantial improvement with respect to the reported experiments 
by injecting a coherent driving in the harmonic mode. Besides, 
the output power is highly enhanced.

\acknowledgments
C.\ C.\ thanks S.\ Schiller and specially A.\ G.\ White for useful comments 
and suggestions. Work supported in part by grants No.\ TIC95-0563-C05-03, 
No.\ PB96-00819, CICYT (Spain) and Comunidad de Madrid 06T/039/96 
(Spain).

\end{document}